\documentstyle[12pt,aasms4]{article}
\def\rosat{{\sl ROSAT~}}
\def\asca{{\sl ASCA~}}

\begin{document}

\title{A2111: A $z=0.23$ Butcher-Oemler Cluster with a Non-isothermal 
Atmosphere and Normal Metallicity}

\author{Mark Henriksen}
\affil{Physics Department, University of North Dakota,
    Grand Forks, ND 58202-7129}
\centerline{mahenrik@plains.NoDak.edu}
\author{Q. Daniel Wang}
\affil{Dearborn Observatory, Northwestern University, 2131 Sheridan Road, Evanston, IL 60208-2900}
\centerline{wqd@nwu.edu}
\author{Melville Ulmer}
\affil{Department of Physics and Astronomy, Northwestern University, 2131 Sheridan Road, Evanston, IL 60208-2900}
\centerline{m.ulmer2@nwu.edu}

\begin{abstract}
We report results from an X-ray study of the Abell 2111 galaxy cluster 
using the Advanced Satellite for
Astrophysics and Cosmology ({\it ASCA}) and the {\it ROSAT} Position Sensitive 
Proportional Counter (PSPC). By correcting for the energy-dependent 
point-spread function of the {\it ASCA} instruments, we have examined the temperature 
structure of the cluster. The cluster's core within 3' is found to have a
temperature of 6.46$\pm$0.87 keV, significantly higher than 3.10$\pm$1.19 keV in
the surrounding region of r = 3 - 6'. This radially decreasing temperature 
structure can be parameterized by a polytropic index of $\gamma$ $\simeq$ 1.45. 
The X-ray morphology of the cluster appears elongated and clumpy on scales 
$\le$1'. These results, together with 
earlier {\it ROSAT} and optical studies which revealed that the X-ray centroid 
and ellipticity of A2111 shift with spatial scale, are consistent with the
hypothesis that the 
cluster is a dynamically young system. Most likely, the cluster has
recently undergone a merger, which may also be responsible for the 
high fraction of blue galaxies observed in the cluster. Alternatively,
the temperature structure may also be due to the gravitational potential
of the cluster. We have further measured 
the emission weighted abundance of the X-ray-emitting intracluster medium as 0.25$\pm$0.14 solar. 
This value is similar to those of nearby clusters which do not show a large blue 
galaxy fraction, indicating that star formation in disk galaxies and subsequent 
loss to the medium do not drastically alter the average abundance of a 
cluster. This is consistent with recent results which indicate that 
cluster abundances have remained constant since at least z $\sim$0.3. 

\end{abstract}

\section{Introduction}

Both optical (Geller \& Beers 1982) and X-ray (Jones \& Forman 1992)
morphological studies of galaxy clusters indicate
that a significant fraction of nearby clusters have substructures that are possibly due to 
mergers. Temperature maps derived from spectro-spatial X-ray observations
	are a necessary complement to the X-ray, optical, and radio imaging data 
	in the sense that hydro-dynamical simulations of subcluster mergers show that
	heating of the cluster atmosphere may be 
	present in a recent post-merger
	system even when evidence of a merger is not visible in the X-ray surface 
	brightness morphology (Evrard, Metzler, \& Navarro 1996). On the other hand,
	there are clusters such as A2256 which exhibit structure in all
	three wavebands consistent with a merger yet the temperature map obtained 
	with {\it ASCA}
	indicates a quiescent dynamical state (Markevitch 1996).
Spatially resolved 
spectroscopy can thus help us to find hot spots similar to those seen in the 
simulations (Roettiger, Loken, \& Burns 1997; Evrard, Metzler, \& Navarro 1996), which
together with the optical, X-ray imaging, and radio observations, provide a
detailed description of the dynamical state of the cluster.
Such spectral analysis has been carried out for a number of clusters with data from
{\it ASCA}, which has a broad energy coverage and modest spatial 
resolution. While the X-ray spectroscopic evidence of merger may be 
difficult to obtain for some clusters (e.g., A2256) in
others the asymmetric 
X-ray morphology and temperature structure are consistent with those
seen in simulations of subcluster merger.
Such examples are A754 (Henriksen \& Markevitch 1996),
the Coma cluster (Honda et al. 1997), and A1367 (Donnelly et al. 1998).

As violent events, subcluster mergers may also affect the evolution of galaxies. 
Relevant processes include ram-pressure (White et al. 1991), the tidal effect 
from the cluster potential (Henriksen \& Byrd 1996), and ``galaxy
harassment" (Oemler, Dressler, \& Butcher 1997). Consequently, 
properties of cluster galaxies may be intimately connected to the changing 
dynamical state and galaxy environment of clusters (e.g., Kauffmann 1995; 
Oemler, Dressler, \& Butcher 1997). Clusters at early epochs ($z \gtrsim 0.2$)
tend to contain higher fractions of blue galaxies ---
the Butcher-Oemler effect. {\it HST} observations indicate that the
effect results from a high rate of star formation in spiral galaxies
(Dressler et al. 1994; Couch et al. 1994) and from a high fraction
of disturbed galaxy systems (Oemler, Dressler, \& Butcher 1997).  
Based on a study of 10 Butcher-Oemler clusters, Wang \& Ulmer (1997)
have revealed a correlation between the blue galaxy fraction and the X-ray 
isophote ellipticity. A2111 at $z = 0.23$ is one of the clusters in the sample 
and contains a high fraction of blue galaxies ($f_b = 0.16$). Based on 
{\it ROSAT} PSPC and HRI observations, Wang, Ulmer, \& Lavery (1997; hereafter 
WUL) have further reported that A2111 has a highly asymmetric
X-ray morphology and the X-ray centroid and ellipticity shift with 
spatial scale, which suggests that the cluster may be undergoing a merger.

In this paper, we present a spatial-spectral analysis, using an 
{\it ASCA} observation, complemented by the {\it ROSAT} PSPC data
of A2111. This analysis enables us to search for spatial and 
spectral signatures of a merger over a broad energy band 
and to compare the metal abundance of A2111 with nearby clusters. 
Throughout the paper, H$_{0}$ = 50 km sec$^{-1}$ Mpc$^{-1}$ is used,
and 90\% confidence error bars are quoted on all quantities.

\section{Observations and Analysis}

A2111 was observed on January 15-16, 1997 with {\it ASCA} 
for 30,000 seconds. Data was obtained with both the GIS and SIS; 
each has two sensors. The GIS has a higher effective area 
at higher energies ($>$ 5 keV) than the SIS so that use of all
4 data sets is optimum for studies of multi-component emission. 
The data were filtered using
the REV2 criteria utilized by the {\it ASCA} Data Processing
Center. Data were excluded under the 
following conditions: with a radiation belt monitor (RBM) count $>$ 100 cts/s, 
during earth occultation or at low elevation angle to the Earth ($<$ 5 degrees 
for the GIS and $<$ 10 degrees for the SIS), when the
pointing was not stable (deviation of $>$ 0.01 degrees), 
during South Atlantic Anomaly passage, and when the cutoff rigidity (COR) was
$>$ 6 GeV/c. Additionally, the SIS was required to be $>$ 20 degrees
to the bright earth and were cleaned to remove
hot pixels. The resulting good exposure times are given in Table 1
The {\it ROSAT} PSPC observations have been discussed in WUL. Briefly,
the observations have an exposure of 7511s, a spatial resolution
of $\sim$0.5', and about 7 overlapping energy bands in the 0.1-2 keV range.

To obtain an emission weighted spectrum for the cluster, we first 
conducted a joint fit to the spectra from the {\it ROSAT} PSPC and the 
{\it ASCA} GIS and SIS detectors. Extracted from a region
within 6' from the assumed cluster centroid at 
$15^h39^m 36^s.554; +34^\circ 25^\prime 31''.16$ (R.A.; Dec.; J2000),
the spectra include essentially all of the cluster emission. 
Background for the PSPC, taken from source-free regions of the image,
is calculated from 4 circular regions of radius 7.3' located at:
(15:40:56.408, +34:50:13.32), (15:36:41.344, +34:27:59.74),
(15.38:16.532, +33:51:05.97), and (15:42:17.971, +34:10:45.85).
The SIS data
was taken in 1-ccd mode and the cluster essentially fills the chip, 
we thus utilized blank sky, deep {\it ASCA} observations taken at high Galactic 
latitudes for background subtraction. The GIS background
was extracted  similarly to avoid uncertainties related to 
vignetting, shadowing of the instrument supports, and gain variations
with radius from the detector center.The energy bands used are: 0.1 - 2 keV
for the PSPC, 0.3 - 10 keV for the SIS, and 0.7 - 10 keV for the
GIS. We adopted the Raymond \& Smith thermal 
plasma model. The two GIS normalizations were
fixed to have the same emission integral, as were the two SIS normalizations. 
The redshift of the cluster was taken to be 0.23.
The abundance, column density, and temperature were left as free parameters 
giving a total of 5 free parameters. We fit this model to the 2 GIS
data sets with free normalizations and found that the normalizations
were essentially identical, as expected. This test was repeated
for the 2 SIS data sets yielding the same result justifying
tying the normalizations as described above. 
The fit to all 5 data sets is not acceptable with a reduced 
$\chi^{2} = 344.2$ for 311 degrees of freedom. The data and the best 
model fit are presented in the top panel of Fig. 1 and the
residuals are shown in the bottom panel.

	While the above analysis was not sensitive to any temperature structure
in A2111, we measured the ICM temperature of the cluster in two 
regions, with radii 0-3' and 3-6'. Further dividing the regions was not practical
due to  the limited extent of the cluster compared to the
XRT+GIS PSF and the limited counting statistics of the \asca observation.
The temperature measurement used a PSF modeling technique
described in Markevitch (1996) and Takahashi et al. (1995).  
This technique has been successfully used in similar analyses for several 
relatively low redshift clusters (see references in Markevitch,
Sarazin, \& Henriksen 1997). Briefly, the PSPC image is used
as a model surface brightness template which is convolved
with the \asca mirror effective area and PSF to produce
model spectra in the two regions. The {\it ROSAT} image was
flat fielded, background subtracted, and rotated to match the
GIS roll angle. The PSPC energy range used is 0.5 - 2.0 keV
and the emission measure is corrected to the {\it ASCA} energy
band. Since the PSPC is used as a surface
brightness template to get  the EM for the spectral fits, a slightly
higher band (0.5 - 2 keV) was used to better match the GIS and SIS.
Channels for each data set are grouped
to contain at least 20 counts. The model PSF, which is based on
GIS observations of Cyg X-1 at various radii from the
detector center (Takahashi et al. 1995), is increasingly
uncertain at low energies so the minimum energy used was 1.5 keV
to minimize this uncertainty. To maintain greater than approximately
20 counts in any fitted energy bin, the data were grouped to give
energy bins of 1.5-2.5, 2.5-4., 4.-7. keV in the SIS and 1.5-2.5, 2.5-3.,
3.-5., 5.-7., 7.-11 keV in the GIS. 
Markevitch (1996) discussed in detail various consistency checks 
performed in validating the use of the method for {\it ASCA} data.

Since the A2111 observation and the deep, blank sky 
observations used in the background subtraction were taken at different times, 
their COR values are different. We thus subtracted the background
using blank sky images, each at a specific COR value, weighted
to the amount of source data obtained at that COR value.
The SIS background image was normalized
by exposure time. A 20\% systematic error in the 
SIS and a 5\% error in the GIS background
normalization were included in the fitting procedure.
The SIS error was estimated at 20\% based on
a day-to-day variation in the GIS background of $\sim$20\%
for a specific COR value.
By using a composite background consisting of
GIS observations with the same COR values as the
data, the
GIS background was better determined and the
error in the normalization was estimated at 5\% (Markevitch
1996 and references within). There errors were then added in quadrature
with the random errors. Table 1 presents the resulting number of background
subtracted counts in each of the regions from each detector integrated
over the
full energy band.

We simultaneously fitted the four spectra from each region using the
Raymond \& Smith model while fixing the
abundance at the best fit value from the single region fit, 0.25 Solar.
We fixed the column density at the measured 21 cm value, 1.9$\times$10$^{20}$
cm$^{-2}$, because the data used, $>$1.5 keV,
are insensitive to the exact value.

Confidence intervals on temperature were estimated by the following procedure. 
A Monte-Carlo simulation of the number of counts in each energy
band of the spectra, assuming a Gaussian distribution of counts
around the observed value, was carried out and the spectra were fitted to obtain 
the best-fit temperature. 
Two hundred simulated spectra were fit and the variance
of the distribution of best-fitting temperatures was calculated to
obtain the 90\% confidence range. A systematic error of 5\% each 
for the PSF and effective area are included in the error simulation.

The best fit models and GIS2 data are
presented in Figs. 2 and 3 for the 0-3' and 3-6' regions respectively.

\section{Results and Discussion}

We present in Figs. 4 and 5 the exposure-corrected SIS and GIS contour maps. 
Both maps show an overall elongation of the cluster X-ray morphology. The X-ray
intensity distribution in the SIS map is very clumpy, compared to that in 
the GIS map. Similar features also appear in the PSPC data. The statistical significance 
of {\sl individual} features is marginal, however.
But the overall clumpiness of the X-ray distribution is real, since the SIS and
GIS maps were smoothed in the exactly same way to have the same noise level.
The PSF of the ASCA X-ray telescopes (XRT) has a relatively sharp core 
(FWHM of  $\sim$50 arcsec) but broad, energy dependent wings 
which extend to a half-power diameter of 3 arcmin. 
The intrinsic spatial broadening of the SIS is negligible compared to that of 
the XRT. The GIS has its own PSF characterized by broad low energy wings
which adds to the XRT PSF. Thus, the clumpy structure appears much more clearly 
in the SIS map. The clumpy X-ray morphology may arise from the presence of
multiple components of the ICM. Assuming an approximate pressure balance,
the temperature inhomogeneity could naturally result in large emission measure
differences in the ICM, which is manifested in the X-ray emission.

The results from our spectral modeling are summarized in Table 2.
The emission weighted temperature for the cluster derived
from a joint fit of the {\it ROSAT} and {\it ASCA} data is 4.9 - 5.9 keV
(90\% limit), which overlaps the results reported by WUL based solely on the PSPC
(2.1 - 5.3 keV). The 90\% confidence range on the column density, 
1.03 - 1.36$\times$10$^{20}$ cm$^{-2}$ is slightly below that
measured from 21 cm, 1.9$\times$10$^{20}$ cm$^{-2}$. However, refitting the
model with the column density fixed at the Galactic value
gives a $\chi^2$ is 378.2/312 degrees of freedom, an increase
in $\chi^2$ of 33.4. The preference
for a column density below Galactic for the single temperature component model
may be due to the overall poor fit of an isothermal model.

Our spatial-spectral analysis of A2111 further suggests that the average 
temperature of 
6.46$\pm$0.87 keV in the central region ($r < 1$~Mpc) of A2111 is significantly 
higher than 3.10$\pm$1.19 keV in its surrounding, r = 3 - 6'. A higher temperature
in the central region
is consistent with the results from simulations after a subcluster has 
passed through the core of the main cluster (Roettiger, Loken,
\& Burns 1993), supporting
the hypothesis that A2111 is undergoing a merger.
Using only the PSPC data, WUL found
     that if the column density is fixed at the
     Galactic value, the subcomponent has a higher temperature
     than the rest of the cluster. This is consistent with the heating in the {\it ASCA}
     temperature map.  Thus, A2111 is the
the first intermediate redshift cluster which 
contains a large blue galaxy fraction for which the optical, X-ray imaging,
and X-ray spectral data are all consistent with the interpretation
of a merger. 

As a test of the robustness of the {\it ASCA} 
PSF correction, Donnelly et al. (1998) applied two independent methods of correcting
for the {\it ASCA} 
PSF (one of which was used for this paper) 
in analyzing the A1367 data. Similar features in the derived temperature
maps were obtained using each methods.

	The only cluster with a
similar redshift for which a similar spatial-spectral
study using {\it ASCA} data has been conducted, is the
z = 0.2 cluster, A2163 (Markevitch
1996). In A2163, the temperature drops with radius out to 3 Mpc, consistent
with a polytropic index ($\gamma$)of 1.9. The cluster atmosphere is apparently 
convectively unstable after a very recent major merger.
The temperature profile for A2111 is less steep than A2163, the equivalent
$\gamma$ is = 1.45 (using the density parameters from WUL: $\beta$ = 0.54
and core radius = 0.21 Mpc); perhaps this cluster has passed the
stage of convective instability or involves less massive subclusters. 
Alternatively, a temperature drop with radius may reflect 
a more centrally concentrated gravitational
potential of the cluster rather than shock heating from 
a merger. However, the case for A2111 being
a merger candidate is strengthened by the spectral and spatial
results taken together. 



The best fit abundance of A2111, 0.11 - 0.39 solar (90\% confidence),
is typical of low z clusters and is 
consistent with the studies of large samples (Allen \& Fabian 1998;
Mushotzky \& Loewenstein 1997) 
which indicate that metallicity in
galaxy clusters in essentially constant out to z $\sim$0.3. 
A2111 is unlike the nearby clusters which show a similar abundance
because it has a high frequency of star forming galaxies, while
the nearby clusters do not. While increased star formation will increase
the metallicity of the interstellar medium and subsequent 
enrich the intergalactic medium by a variety od processes, including
ram-pressure and tidal stripping, the similarity in abundance argues against 
this eposodic star formation having a significant effect on the
overall metallicity
of the cluster gas. 

In conclusion, our ASCA data show that A2111 has an elongated and clumpy 
X-ray morphology as well as  a relatively high temperature 
core, compared to the surrounding regions. These results, together with the 
apparent substructure observed in the {\it ASCA} and \rosat observations, 
strongly suggest that
the cluster is undergoing a merger, which is likely responsible for the observed 
large blue galaxy fraction of the cluster. Future X-ray observations will be
necessary to study the relationship between possible element abundance 
gradients in the intergalactic
medium and the blue galaxy distribution in the cluster.

\acknowledgments

MJH thanks the National Science Foundation for support through Grant No. AST-9624716, 
and QDW acknowledge the support from NASA for his research.

\clearpage

\begin{deluxetable}{ccccc}
\footnotesize
\tablewidth{0pt}
\tablecaption{ASCA Data}
\tablehead{
\colhead{Detector}& \colhead{Region} & \colhead{Counts}&
 \colhead{Exposure (sec)} 
}
\startdata
GIS 2 & 0-3' & 644.5 & 23707. \nl
  & 3-6' & 402.8 &  \nl
GIS 3 & 0-3' & 533.4 & 23370. \nl
  & 3-6' & 335.9 &  \nl
SIS 0 & 0-3' & 494.0 & 13562. \nl
  & 3-6' & 294.0 &  \nl
SIS 1 & 0-3' & 350.4& 14314. \nl
  & 3-6' & 217.6 &  \nl

\enddata
\end{deluxetable}

\begin{deluxetable}{cccccc}
\footnotesize
\tablewidth{0pt}
\tablecaption{Spectral Modeling Results}
\tablehead{
\colhead{Region(s)} & \colhead{$\chi^2$/dof}&
 \colhead{kT(keV)} & \colhead{$n_{H}$cm$^{-2}$} & \colhead{Abundance}
}
\startdata
 0-6' & 344.8/311 & 5.38 +0.50/-0.47 & 1.19 +0.17/-0.16$\times$10$^{20}$  & 0.25 +0.14/0.14 \nl
 0-3'& 8.6/17 & 6.46$\pm$0.87 & - & - \nl
			 3-6'& 9.4/17		& 3.10$\pm$1.19 & - & - \nl
\enddata
\end{deluxetable}
\clearpage
\figcaption[]{The data for the PSPC, GIS2, GIS3,
SIS0, and SIS1 and best fit single component Raymond
and Smith model are shown in the upper panel. 
The reduced $\chi^{2}$ is 1.11. $\chi^{2}$ 
vs. energy is plotted in the lower panel with the
sign of the residual.
}\label{fig1}
\figcaption[]{The best fit model, background subtracted counts and
error are shown in each of the energy bins fit for the G2 in the
inner, 0-3' region of the cluster. The triangles are model counts while the data is shown as a histogram 
with 1$\sigma$ error bars. The errors are the statistical and
systematic errors (effective area, PSF, and background
normalization) added in quadrature. This is representative of the
data quality and the goodness of fit
since the spectrum from the G3, SIS0, and SIS2 were
also fit but are not shown.}\label{fig2}
\figcaption[]{The same as is shown in figure 2 for
the outer, 3 - 6' region of the cluster.}\label{fig3}
\figcaption[]{The exposure corrected 
SIS intensity map is adaptively smoothed with
a Gaussian function, adjusted to achieve a uniform signal-to-noise
of 6. The contour levels are 2$\sigma$ higher than the next lower one
and have values of: 1.6, 2.1, 2.8, 3.8, and 5.0$\times$
10$^{-3}$ counts arcmin$^{-2}$ sec$^{-1}$. The background
level is 1.2$\times$10$^{-3}$ counts arcmin$^{-2}$ sec$^{-1}$.}\label{fig4}
\figcaption[]{The GIS map is prepared similarly to the SIS
map in figure 4. The contour levels 
have values of: 2.0, 2.7, 3.6, 4.7, 6.3, 8.4, 11.2, 14.9,
and 20.0$\times$
10$^{-4}$ counts arcmin$^{-2}$ sec$^{-1}$.
The background
level is 1.5$\times$10$^{-4}$ counts arcmin$^{-2}$ 
sec$^{-1}$.}\label{fig5}
\clearpage
\plotone{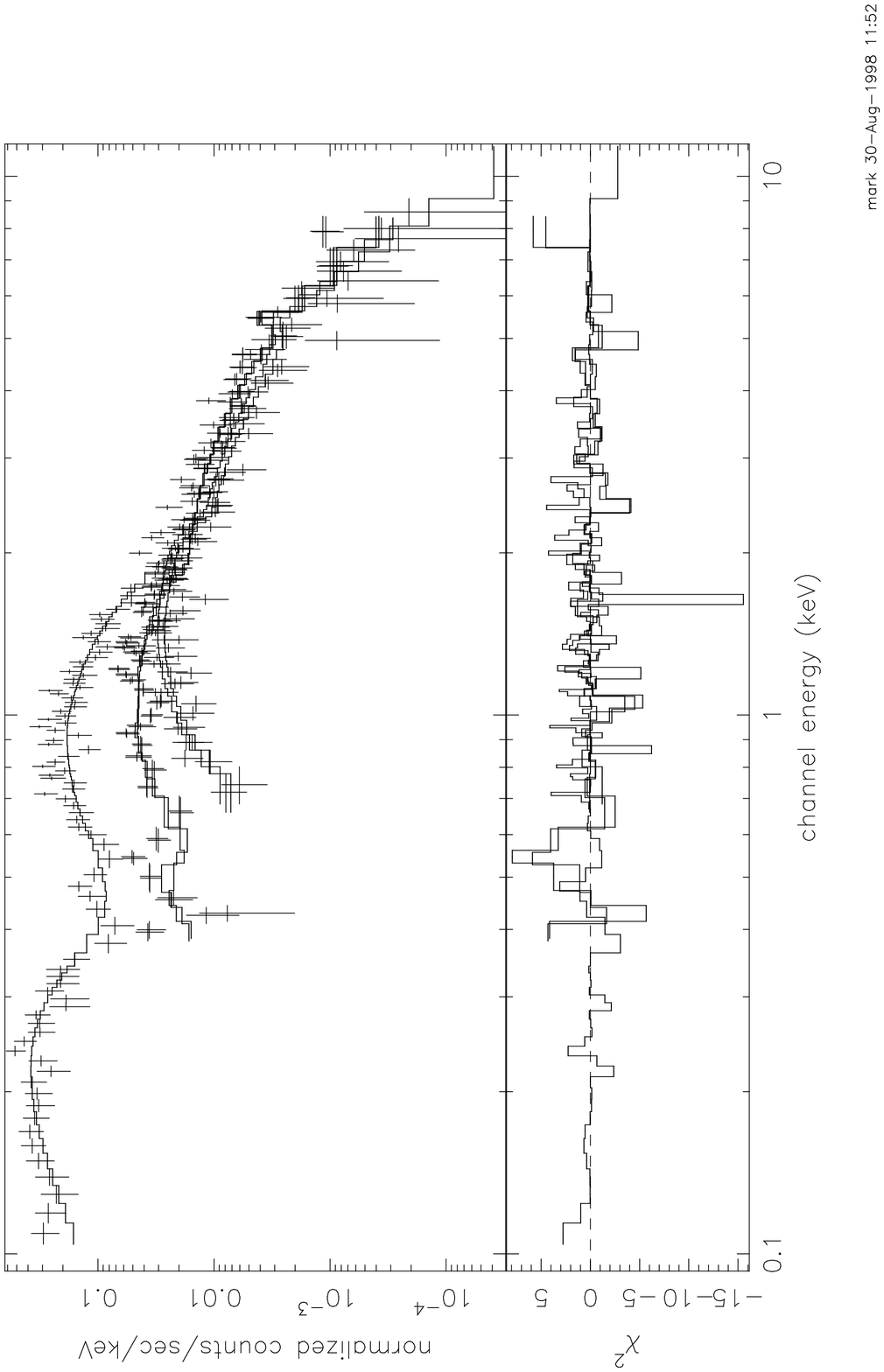}
\clearpage
\plotone{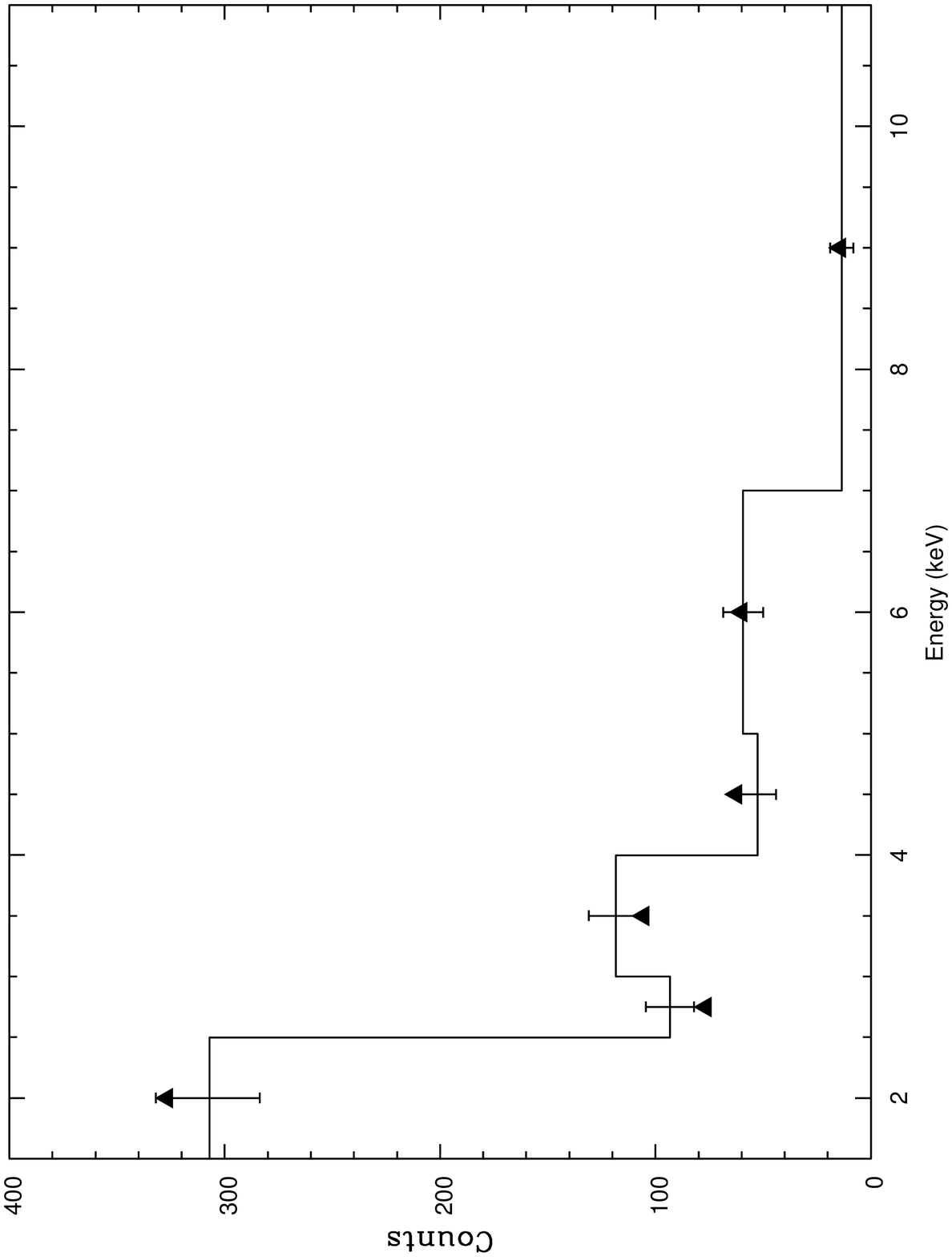}
\clearpage

\plotone{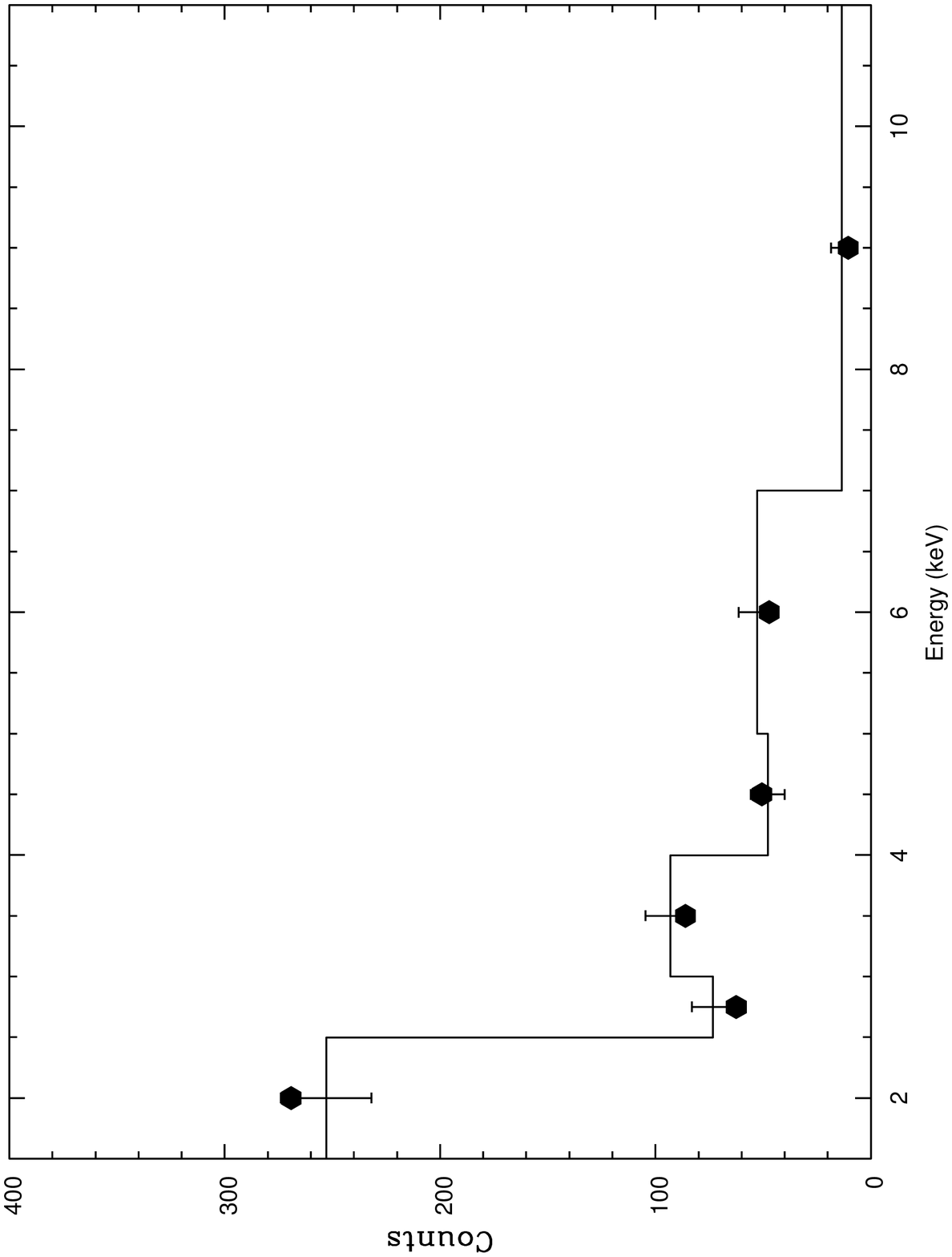}

\clearpage
\plotone{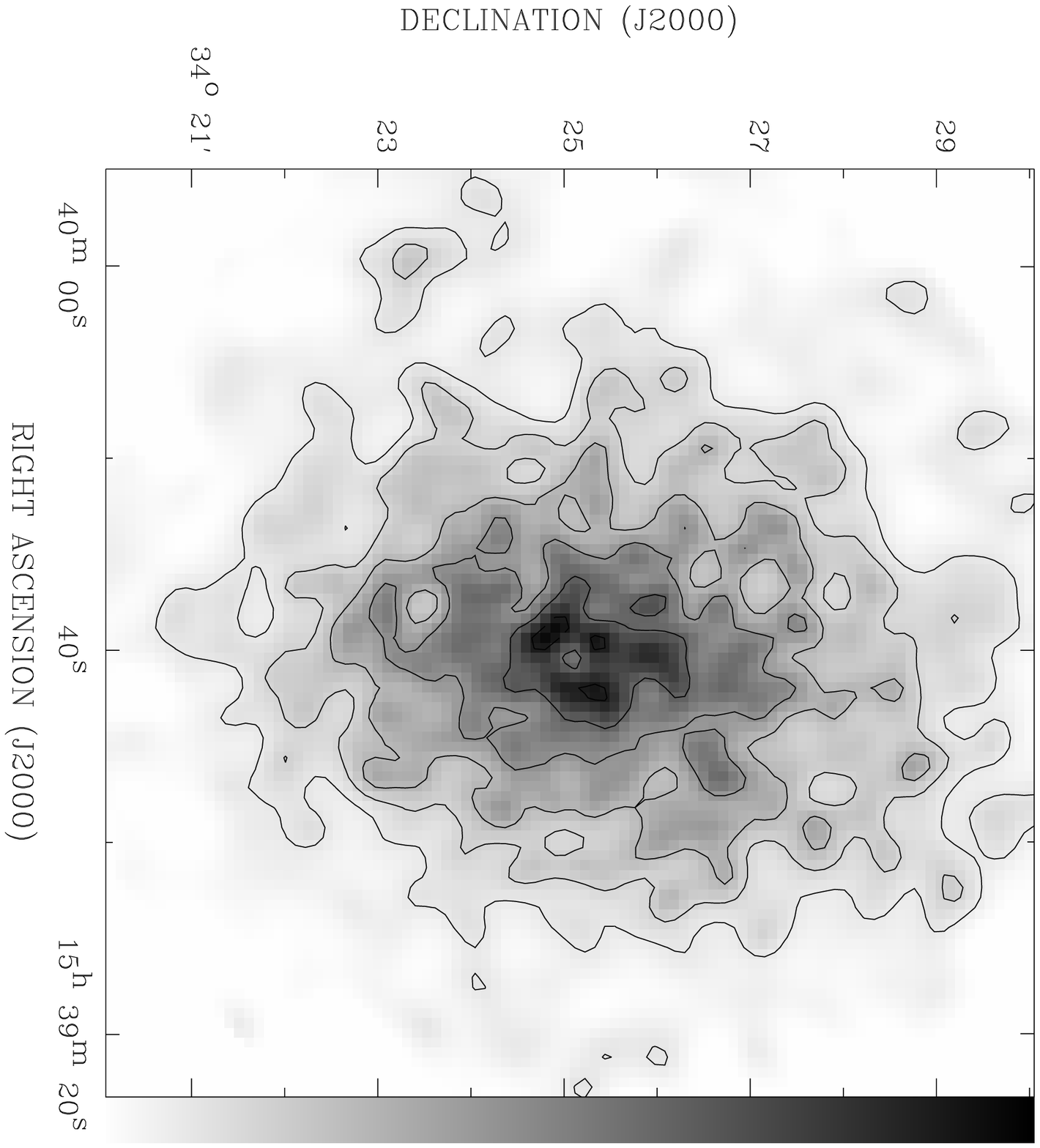}
\clearpage
\plotone{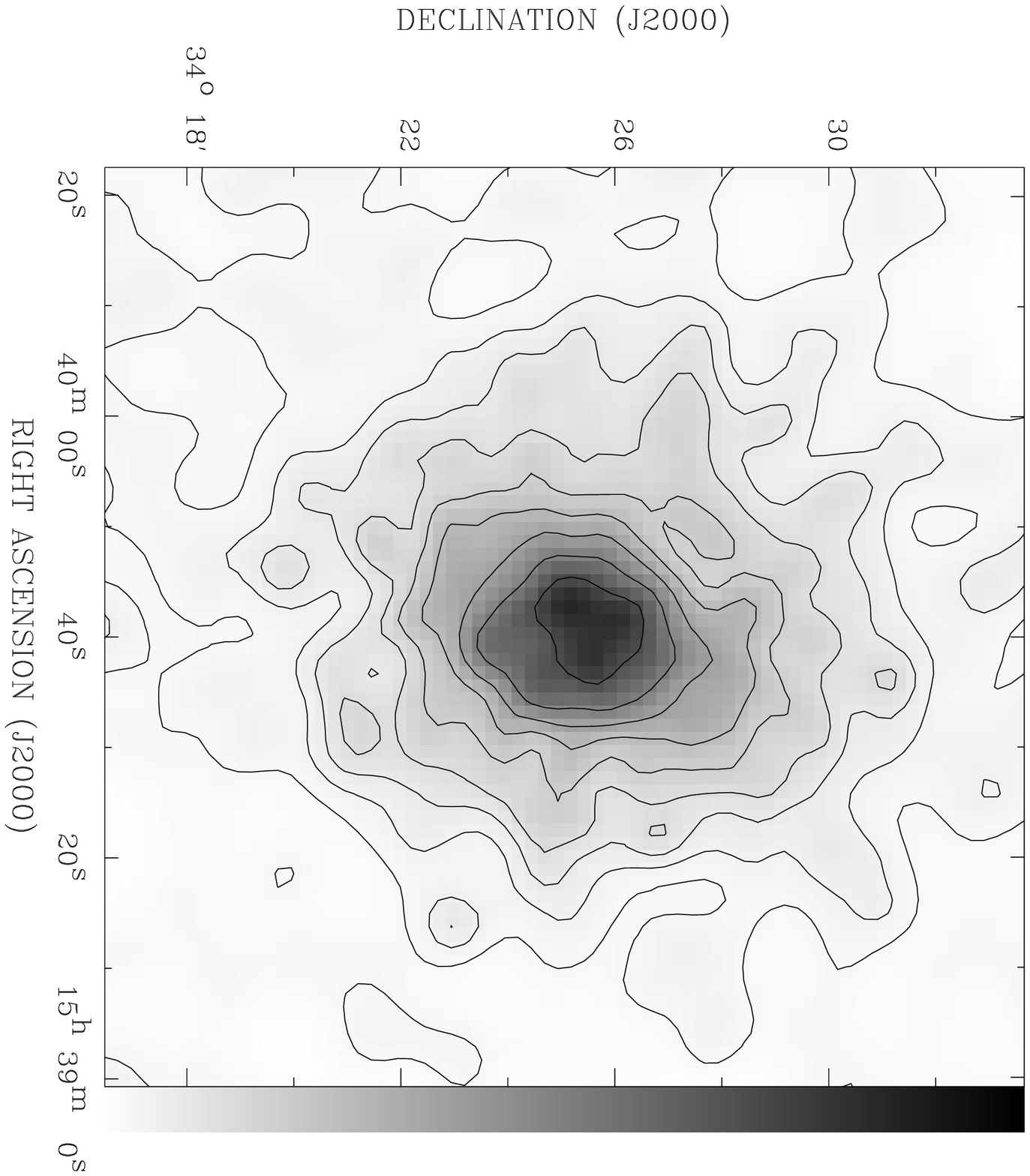}

\end{document}